\documentclass{aa}
\usepackage{graphicx}
\usepackage{txfonts}
\begin{document}
%
   \title{New membership determination and proper motions of
NGC~1817. Parametric and non-parametric approach.
\thanks{Table~5 is only available in electronic form from CDS via
 anonymous ftp to cdsarc.u-strasbg.fr (130.79.128.5) or 
via http://cdsweb.u-strasbg.fr/cgi-bin/qcat?J/A+A/
Fig~1 and Table~1, 2 and 4 are only available in electronic form
via http://www.edpsciencies.org}
    }

   \author{L. Ba\-la\-guer-\-N\'u\-\~nez\inst{1,2,3}, 
   C. Jor\-di\inst{1,4}, 
   D. Galad\'{\i}-Enr\'{\i}quez\inst{5},
   J.L. Zhao\inst{2}
          }

   \offprints{Balaguer-N\'u\~nez, L., \email{Lola.Balaguer@am.ub.es}}

   \institute{Departament d'Astronomia i Meteorologia, Universitat de
            Barcelona, Avda. Diagonal 647.  E-08028 Barcelona, Spain 
         \and
            Shanghai Astronomical Observatory, CAS Shanghai 200030,
            P.R. China
         \and
            Institute of Astronomy, Madingley Road,
	    CB3 OHA Cambridge, UK
         \and
            CER for Astrophysics, Particle Physics and Cosmology,
	    associated with Instituto de Ciencias del Espacio-CSIC
         \and
            Centro de Astrobiolog\'{\i}a (CSIC-INTA), 
            Carretera de Ajalvir km 4, E-28850 
            Torrej\'on de Ardoz, Madrid, Spain 
             }
\date{Received ; accepted}

\authorrunning{Balaguer-N\'u\~nez et al}
\titlerunning{Membership of NGC~1817}

\abstract{
We have calculated proper motions and re-evaluated the membership 
probabilities of 810 stars in the area of two NGC objects, 
NGC~1817 and NGC~1807. We have obtained 
absolute proper motions from 25 plates in the reference system of 
the Tycho-2 Catalogue. 
   The plates
   have a maximum epoch difference of 81 years; and they were taken
   with the double astrograph at Z\^ o-S\`e station of Shanghai
   Observatory, which has an aperture of 40 cm
   and a plate scale of 30\arcsec~mm$^{-1}$. The average proper motion
   precision is 1.55 mas~yr$^{-1}$. These proper motions are used to
   determine the membership probabilities of stars in the region,
based on there being only 
one very extended physical cluster: NGC~1817. 
With that aim, we have applied and compared parametric and 
non-parametric approaches to cluster/field segregation.
We have obtained a list of 169 probable member stars. 
   }

   \maketitle

%

\section{Introduction}

 The open cluster NGC~1817 (C0509+166), in Taurus 
[$\alpha_{2000}$=$5^{\mathrm h}12^{\mathrm m}\llap{.}1$,
$\delta_{2000}=+16{\degr}42\arcmin$], 
is an old and rich but poorly studied open cluster (Friel \cite{Friel}).
NGC~1817 seems to be as old as the Hyades,
with a lower heavy-element abundance. Its location at
1800 pc almost directly towards the Galactic anti-center and 400 pc below the
plane [$l=186^\circ\llap{.}13$, $b=-13^\circ\llap{.}12$] and its metallicity, lower than solar, make it an object 
of special interest for
the research of the structure and chemical evolution of the Galaxy.

   NGC~1807 (C0507+164), also in Taurus [$\alpha_{2000}$=$5^{\mathrm h}10^{\mathrm m}\llap{.}7$,
$\delta_{2000}=+16{\degr}32\arcmin$] shows up as a group of bright stars 
on a mildly populated background,
located close to NGC~1817. 

	A recent determination of mean proper motions of open clusters
(Dias et al.\ \cite{Dias02})
based on the Tycho-2 Catalogue (H\a{o}g et al.\ \cite{tyc2a}) gives absolute values
for NGC~1817. But the study is based on only 19 stars in an area of 15$\arcmin$,
with ten stars considered as cluster members.
The only accurate study of astrometric data, based on automatic 
measurements of 12 plate pairs, (Balaguer-N\'u\~nez et al.\ \cite{Bai}, 
hereafter Paper~I) gave relative proper motions of 722 stars in the area.
These proper motions were analyzed to determine membership for two clusters
in the area: NGC~1817 and NGC~1807.

 	Later on, Balaguer-N\'u\~nez et al.\ (\cite{PaperIII}, 
hereafter referred to as Paper~III) undertook a wide field photometric study 
(1998-2000) of a 65$^{\prime}$$\times$40$^{\prime}$ area around NGC~1817
in the $uvby-H_{\beta}$ system down to a limiting magnitude $V$$\approx22$.
These photometric results confirm that NGC~1807 is not a real 
physical open cluster and that only one very extended open cluster, NGC~1817, 
covers the area. 

Moreover, Mermilliod et al.\ (\cite{Mermi}, hereafter Mer03) 
have determined radial velocities of red giant stars in the area. 
Two stars in the region of NGC~1807 have the same radial velocities as   
the stars in NGC~1817. 

The inappropriate assumption of there being two open clusters in the 
membership analysis of Paper~I, 
could affect the conclusions, as will be shown later. So a new membership 
determination seems advisable and two completely 
different methods have been used for this purpose. This has never been done 
before on the same set of data,
which makes our comparison specially interesting. 

Since the publication of Paper~I, the release of the Tycho-2 Catalogue 
allows an accurate transformation from $x$ and $y$ coordinates derived 
from plate measurements to the ICRS system, leading to proper motions 
computed directly in absolute sky coordinates, which would make the 
resulting catalogue much more useful for further studies. 
Finally, as is shown in this paper, a central overlap 
technique applied to the PDS scan data of all the available 25 plates 
makes it possible to enlarge the sample in Paper~I by   
about one hundred stars. 


    In this paper we determine precise absolute proper motions of
810 stars within a 1\fdg5~$\times$~1\fdg5 area in the NGC~1817 region, 
from automatic PDS measurements of 25 plates. The estimated 
membership probabilities lead us to a complete astrometric study of 
the cluster area.
   Section~2 describes the plate material 
as well as the proper motion reduction procedure and results, with 
comparisons with the Tycho-2 Catalogue. Section~3 
accounts for the membership determination, using parametric
and non-parametric approaches. Section~4 is devoted to 
the analysis of results.
    Finally, a summary is presented in Sect.~5.     

\section{Proper motion reduction}

\subsection{Plate material and measurements}
   Twenty-five plates of the NGC~1817 region are available.
   They were taken with the double astrograph at the Z\^ o-S\`e station
   of Shanghai Observatory. This telescope, built by Gaultier in Paris
   at the beginning of the last century, has an aperture of 40 cm, a focal
   length of 6.9 m and hence a plate scale of 30\arcsec~mm$^{-1}$. The size of the
   plates is 24 cm by 30 cm, or 2\fdg0 $\times$ 2\fdg5.
Although the plates cover quite a wide sky area, only a section of
$1\fdg5\times 1\fdg5$ around the cluster center was measured for this study.
   The oldest
   plate was taken in 1916, and the most recent ones in 1997. 
	The plate material and measurements are basically explained in Paper~I.  
However, as we decided not to use the plate-pairs technique, we include 
detailed information on each individual plate in   
Table~\ref{plates}. 
Moreover, the central overlap technique let us make use of
all of the original 25 Shanghai plates, 
one more than in Paper~I (plate CL82005).

\addtocounter{table}{1}
%
%

\subsection{Proper motions}
    The absolute proper motions in the region of NGC~1817 were obtained
    by following the central overlap procedure
    (Russel \cite{russel}; Wang et al.\ \cite{Wang95}, \cite{Wang96},
    \cite{Wang00}).
    At the time of the previous reduction (Paper~I) 
only 15 PPM (R\"oser \& Bastian \cite{PPM}) stars were available 
as standard stars. So the reduction of relative proper motions
based on plate pairs was the appropiate choice.

    The central overlap method simultaneously determines the plate-to-plate
transformation parameters, the star motions and their errors. 
This method has 
rigorous mathematical foundations (Eichhorn \cite{Eich}), 
but its computational requirements are so huge that, in 
practice, it cannot be implemented in its strict formulation.
The usual approach to the method is generally known as iterative
central-overlap algorithm, and implies the separation of the
determination of plate and star parameters in consecutive steps
that are iterated until convergence is achieved. This procedure
is known to be equivalent, in practice, to the one-step 
block-adjustment approach, 
and has been extensively used during the last decades 
(Wu et al.\ \cite{Wu}, Galad\'{\i}-Enr\'{\i}quez et al.\ \cite{galab}
among others).  
In applying the central overlap technique, the plate measurements
are first reduced to a reference catalogue system, using the data
for those stars with reference positions to determine the plate
constants. 
As initial catalogue, 86 stars from the Tycho-2 Catalogue at epoch 
J2000 (H\a{o}g et al.\ \cite{tyc2a}) were selected.
The plate constants are then applied to all the stellar measurements,
giving equatorial coordinates for each star from each plate on which 
the star appears. The data for each star are then selected and solved by
least squares for improved positions and proper motions. These revised
star parameters form a new catalogue, which is on the system of the
original reference catalogue, but has been strengthened and expanded to 
include additional stars. This new catalogue is used for a new determination
of the plate constants, and the resulting positions solved for a 
further improvement of the stars' positions and proper motions. 

\addtocounter{figure}{1}

The iteration requires equatorial coordinates for each star on 
each plate as starting values. After positions for
each star on each plate are computed, proper motions are determined
from a linear regression of position versus time.
If the error of one proper motion component
exceeded 3$\sigma$, the most deviant measurement was discarded
and the proper motion recomputed, until the error fell bellow the 
mentioned limit.
In the next iteration, all stars with precise proper motions are used 
as reference stars for a new transformation from plate to spherical
coordinates. In the first iteration, only
linear terms were used; in the second iteration, higher order terms 
were included. Only significant transformation terms were kept. 
    To select the best plate constant model, we used 
Eichhorn \& Williams' criterion (Eichhorn \& Williams \cite{Eicwill}, 
Wang et al.\ \cite{Wang82}) and obtained a model with
    six linear constants on coordinates, a
    magnitude and a coma term, and a magnitude distortion term. 
    We require that any star remaining in the final catalogue has 
    at least one measurement 
    from the modern epoch plates. 

    The whole
    process is iterated until convergence is achieved. 
    The criteria for convergence were:
    mean differences in position smaller than 1.1 mas, r.m.s. smaller
    than 3.6 mas and differences in proper motion below 0.1 mas~yr$^{-1}$.
    The final outcome results in 810 stars.

\addtocounter{table}{1}

\begin{figure}
\begin{center}
\resizebox{\hsize}{!}{\includegraphics{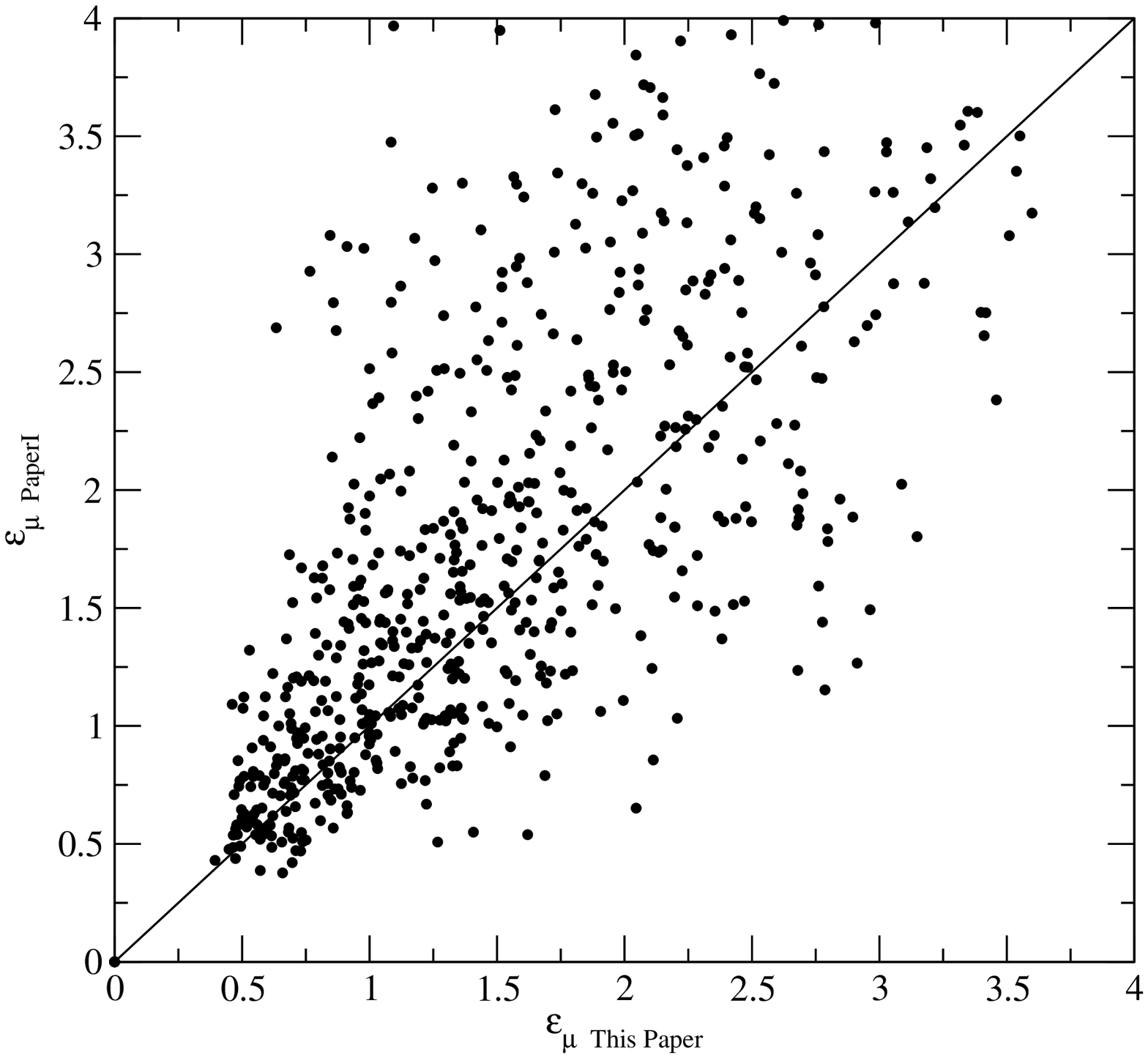}}
\end{center}
\caption{Proper motion errors {\it vs} those in Paper~I.}
\label{errold}
\end{figure}

    Table~\ref{error} shows the mean precisions of final proper motions
    for stars in the NGC~1817 region detected on more than 3 plates.
    21 stars with errors greater than 3 mas~yr$^{-1}$ were not included. 
    Errors for stars on only two plates were not computed.
    The units of the proper motions and their precisions throughout are
     mas~yr$^{-1}$. The precision of the final proper motions strongly depends 
    on the number of plates. 
    Figure~\ref{nplpr} gives the number of stars for which various
    numbers of plates are available. More than $85\%$ of proper
    motions were obtained from at least 5 plates.

\begin{figure}
\begin{center}
\resizebox{\hsize}{!}{\includegraphics{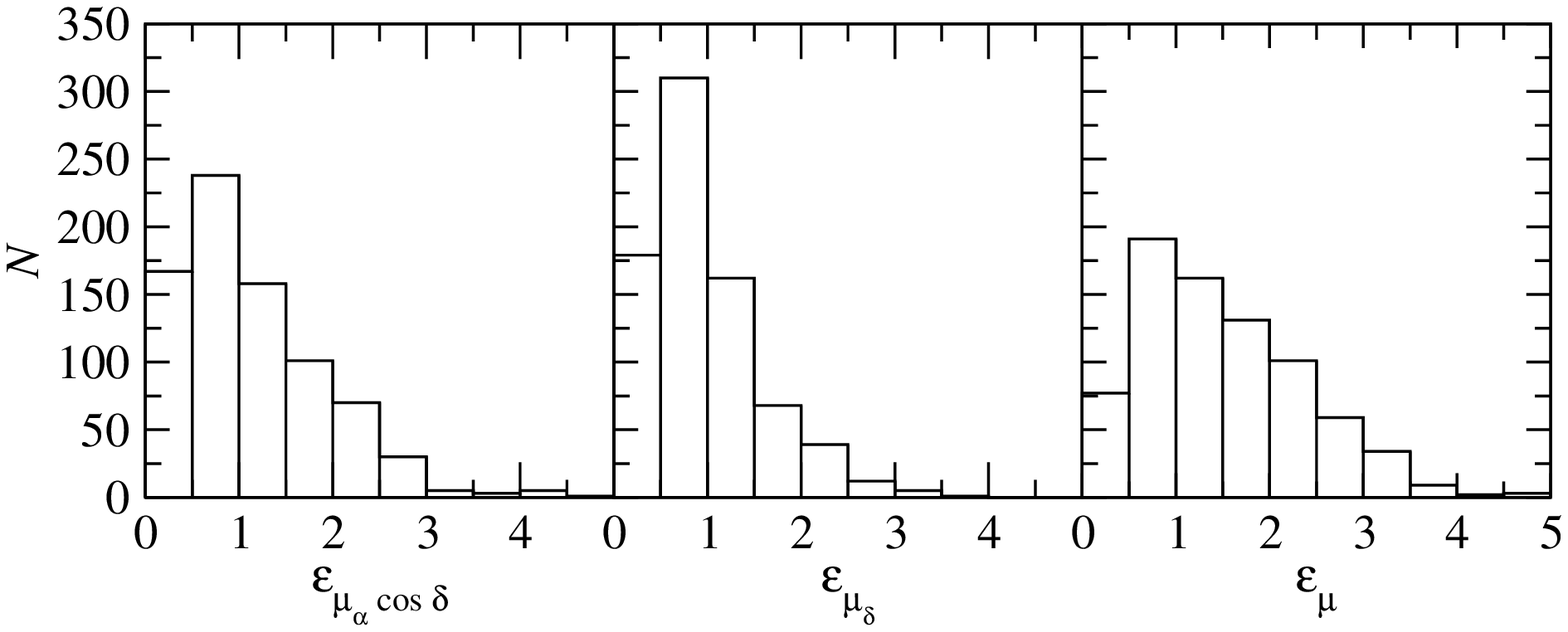}}
\end{center}
\caption{The number of stars {\it vs} the errors in proper motions 
(units are in mas~yr$^{-1}$) }
\label{nerrhis}
\end{figure}

    Thanks to the addition of one more plate and the use of the overlap 
technique we determined proper motions for 88 stars more than in Paper~I. 
Most of these recovered stars were measured in 2-3 plates. In addition, 
the mean number of plates per star also increased by a factor 1.8 (up to 
eight more plates). 
The errors are a factor of 0.77 better than in Paper~I.
Figure~\ref{errold} shows the errors in the proper motions compared
to those in Paper~I.

    The mean errors in the proper motions for more than $80\%$ of the stars
    are ${\epsilon}_{\mu_{\alpha}\cos\delta}$= 1.16 mas~yr$^{-1}$, 
    ${\epsilon}_{\mu_{\delta}}$ = 0.96 mas~yr$^{-1}$
    and ${\epsilon_{\mu}}$= 1.55 mas~yr$^{-1}$, where
    \begin{math}\epsilon_{\mu}=\sqrt{{\epsilon ^{2}_{\mu_{\alpha}\cos\delta}+\epsilon^{2}_{\mu_{\delta}}}} 
    \end{math}. In the most precise case, the errors are 
    0.97 mas~yr$^{-1}$ for stars with more than 21 plates ($32\%$ of stars).
    Figure~\ref{nerrhis} shows the distribution of proper motion errors 
with the number of stars: $N$
    {\it vs\/} $\epsilon _{\mu_{\alpha}\cos\delta}$, $\epsilon _{\mu_{\delta}}$ and
    $\epsilon_{\mu} $.

\begin{figure}
\begin{center}
\resizebox{\hsize}{!}{\includegraphics{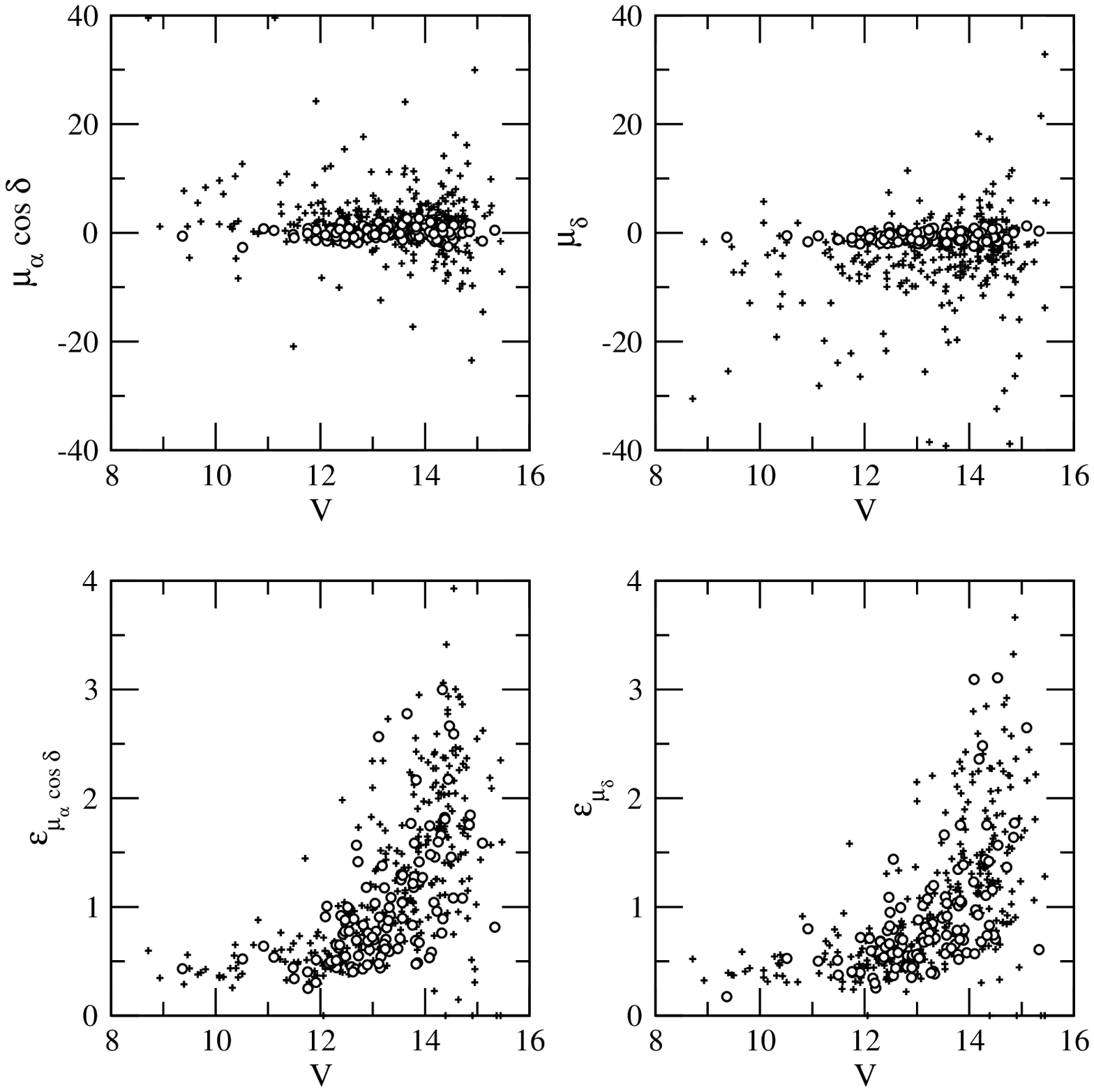}}
\end{center}
\caption{Proper motions (top) and their errors (bottom) {\it vs} $V$ apparent 
magnitude, for the stars
in common with the photometric study (Paper~III). Open circles denote 
selected member stars.
     Null errors are from proper motions calculated with only two plates, (units are mas~yr$^{-1}$).}
\label{mag}
\end{figure}

    Figure~\ref{mag} gives $\mu_{\alpha}\cos\delta$, $\mu_{\delta}$ 
    and their errors as a function of 
    $V$ magnitude of the stars in common with Paper~III. 
    Since the CCD photometry in Paper~III covers a smaller area than
the astrometric catalogue from this paper,
these graphs cannot display the data for all the stars present in this
study, but they describe well the behaviour of the data as a
function of apparent brightness.
    No systematic trends in proper motion are apparent  
    as a function of magnitude for member stars (see Sect.~4). 

\begin{figure}
\begin{center}
\resizebox{\hsize}{!}{\includegraphics{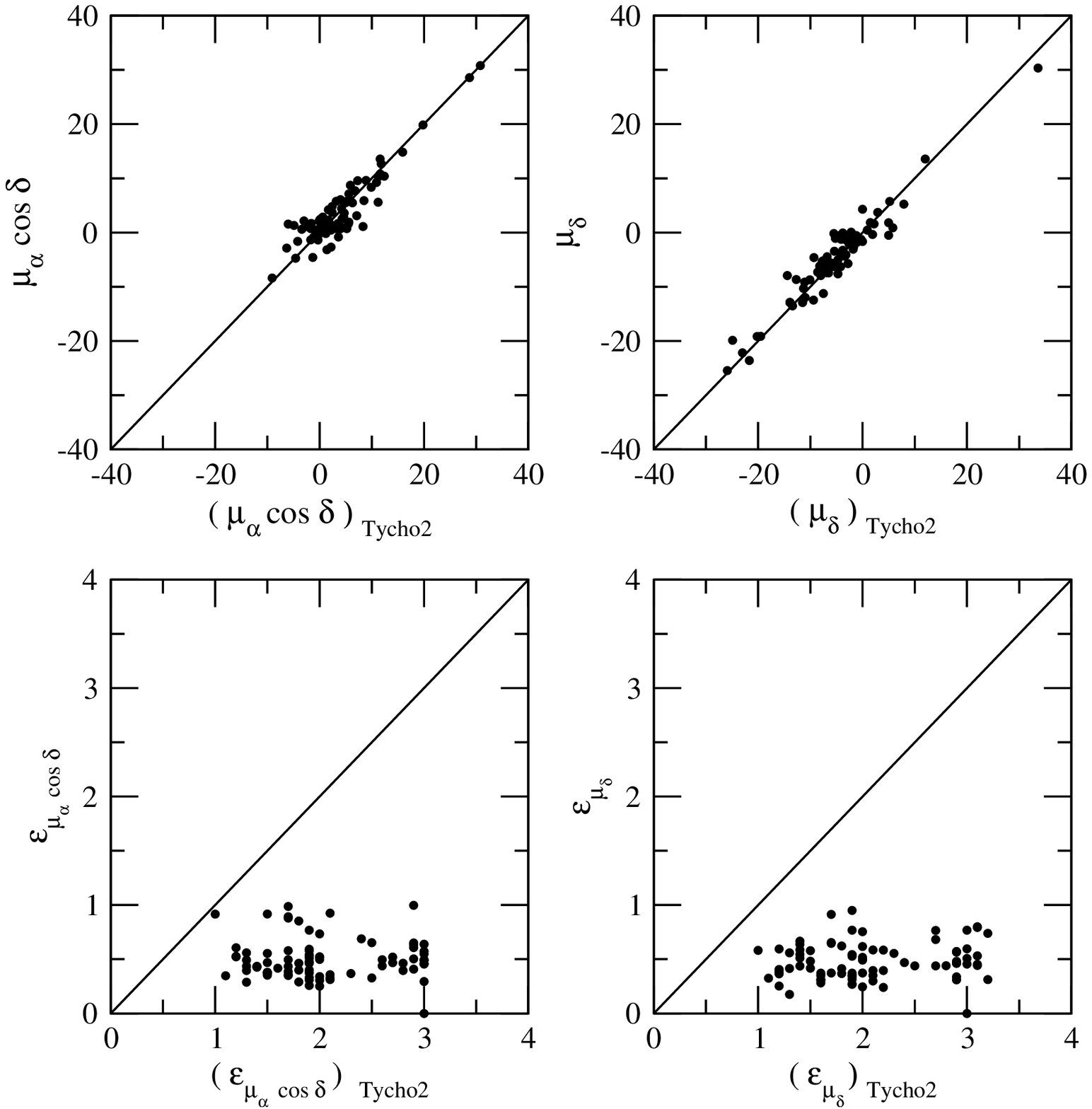}}
\end{center}
\caption{Proper motions and their errors from this paper compared to those 
in Tycho-2 catalogue. (units are mas~yr$^{-1}$)}
\label{comp}
\end{figure}

    Our absolute proper motions and their errors are compared with those of 
    the Tycho-2 Catalogue
    in Fig.~\ref{comp}. 
    Mean differences in the sense ours minus Tycho-2 are $-$0.099 ($\sigma$ = 2.592) 
    and 0.659 ($\sigma$ = 2.557) mas~yr$^{-1}$ in $\mu_{\alpha}\cos\delta$ and $\mu_{\delta}$, respectively.
    A linear fit to the proper motion data gives us:

    $\mu_{\alpha}\cos\delta$ = $-0.010 \ ({\pm} 0.300) + 0.988 \ ({\pm} 0.014)
     \cdot (\mu_{\alpha}\cos\delta)_ {\mathrm{Tyc2}}$ ;\
    $r$ = $0.992$ 

    $\mu_{\delta}$ = $0.406 \ ({\pm} 0.288)  + 0.974 \ ({\pm} 0.010)
     \cdot (\mu_{\delta})_{\mathrm{Tyc2}}$ ;\
    $r$ = $0.995$ 

    \noindent  where $r$ is the correlation coefficient.

\section{Membership determination}

\subsection{The classical approach}
	    Accurate membership determination is essential for
    further astrophysical studies of clusters. The fundamental mathematical
    model for cluster-field segregation set up by Vasilevskis et al.\ (\cite{vas}) 
and the technique
    based upon the maximum likelihood principle developed by Sanders
    (\cite{san}) have since been continuously refined.

    An improved method for membership determination of stellar
    clusters based on proper motions with different observed precisions was
    developed by Stetson (\cite{ste}) and Zhao \& He (\cite{zhaohe}).
	This model has been frequently used
    (Wang et al.\ \cite{Wang95}, \cite{Wang96}, \cite{Wang00} among others).

    We used a maximum likelihood method with a 9-parametric
gaussian model for the frequency function, as described in Wu et al.\ 
(\cite{Wu}).
The quality of the fit should be optimized near the
vector point diagram (VPD) region occupied by the cluster stars, 
where the model is most crucial for
providing reliable membership determinations. Outliers cause the estimated
distribution of field stars to be flatter than the actual one, thus increasing
the final probability of cluster membership (Kozhurina-Platais et al.\  \cite{kozhu},
Cabrera-Ca\~no \& Alfaro \cite{canno}, Zhao \& Tian \cite{Zhao},
Zhao et al.\ \cite{Zhao82}).
To minimize the effect of high proper-motion field stars in the model,
we restricted the membership determination to the range $|\mu| <$  30 mas~yr$^{-1}$.

By applying the standard maximum likelihood method, several 
drawbacks of the parametric method were identified. 
The cluster motion is not very well separated from the mean field motion,
making the convergence of the results unstable and producing an unrealistic 
intrinsic velocity dispersion of the cluster:    
the circular gaussian distribution intended to fit the cluster tends to assume an
excessive width to improve the representation of the field distribution.
This does not happen when this method is applied to dense clusters
that stand out on the field, when the field population represents
a small fraction of the total number of stars. But this behaviour seems
unavoidable in situations like ours, with a cluster with very low contrast
with the field. In these cases the parametrization of the field model is
difficult and, at the same time, crucial, because the residuals of an 
inaccurate field model can be of the same
order of or bigger than the cluster signal, and the membership 
calculations can be affected by this.  
As the cluster is at a distance of 1.8 kpc, it is clear that our 
measurements lack the
resolution needed to resolve its internal velocity dispersion.
And thus, to be on the safe side we decided to set the cluster intrinsic
dispersion to zero in order to minimize the problem just outlined. 
This way, the model will assign to the cluster gaussian 
distribution a width related only to the measurement errors. 
We have tried fixing different internal velocity dispersions
for the cluster, in the range of plausible velocities (1 to 3~km~s$^{-1}$) and 
distances (1.5 to 2~kpc), but the 
slight differences do not affect the values obtained or the segregation, 
since the membership probability is a relative quantity.
We obtained
the distribution parameters and their corresponding uncertainties
shown in Table~\ref{para}. 

\begin{table*}
\leavevmode
\caption{Distribution parameters and their uncertainties
 for the NGC~1817 cluster and the field.
 The units of $\mu$ and $\sigma$ are mas~yr$^{-1}$ }
\begin {tabular} {lc c c c c c c c c }
\hline

& $n_c$
& $ \mu_{\alpha}\cos\delta$ &
$\mu_{\delta}$ & $\sigma_{\mu_{\alpha}\cos\delta}$ & $\sigma_{\mu_{\delta}}$ & $\rho$\\
&
&  &  &  &  &  &  &    \\
\hline
NGC~1817*& 0.261 & 0.29 & $-$0.96 &  &  &     \\
& $\pm0.020$ & $\pm0.10$ & $\pm0.07$ &  &  &    \\

field  &
& 2.29 & $-$4.25  & 5.69 & 6.38 & $-0.08$\\
&   & $\pm0.02$ & $\pm0.27$ &$\pm0.02$ & $\pm0.14$ & $\pm$0.03 \\
\hline
* $\sigma_c =$0 (fixed)
\label{para}
\end {tabular}
\end {table*}

\subsection{The non-parametric approach}

As discussed by many authors (Cabrera-Ca\~no \& Alfaro \cite{canno},
for example), 
the membership determination based on fits of parametric probability density 
functions (PDFs) 
has several limitations.
A circular bivariate function is a good representation of the cluster PDF if
the intrinsic velocity dispersion of the cluster is not resolved. 
Moreover, the choice of an elliptic bivariate 
gaussian function for the
field distribution is known to be unrealistic. The proper motion distribution 
of field stars has an intricate structure dominated by the combination of solar motion
and galactic differential rotation. Furthermore, real field distribution wings 
are stronger than those predicted by a gaussian model (Marschall \& Van Altena 
\cite{Mars}). Soubiran (\cite{Sou}) modeled the field population in the direction
of the North Galactic Pole by means of the sum of three gaussian distributions. 
But adding further gaussians to the field in the classical parametric model
has been shown to give poor results (Galad\'{\i}-Enr\'{\i}quez et al.\ \cite{gala2},
hereafter Gal98).
 
In our case, as seen in Paper~I and in the previous section, the cluster 
mean proper motion is close to the maximum of the field distribution and 
the cluster is loosely concentrated, 
making necessary an accurate model of the field distribution. 
 	Following Gal98, we perform an empirical determination of the PDFs 
without relying on any previous assumption about their profiles.
For a sample of $N$ individuals distributed in a two-dimensional space with coordinates
$(a,b)$, it is possible to tabulate the frequency function $\Phi(a,b)$ by evaluating the
observed local density at each node of a grid of $n_a \times n_b$ points 
extending over
the region of interest in the space. If the grid is dense enough, the empirical frequency
function $\{\Psi(a_1,b_j);i=1,...,n_a;j=1,...,n_b\}$ will be equivalent, for all
practical purposes, to the true $\Phi(a,b)$. The kernel used to estimate that local density 
around a point $(a_i,b_j)$ will be a normal circular kernel. The smoothing parameter $h$ 
(gaussian dispersion), is chosen using Silverman's rule (\cite{Sil}). The procedure was
tested for several subsamples applying different proper motion cutoffs. Satisfactory
results are obtained with a proper motion cutoff of $|\mu| \leq$ 15 mas~yr$^{-1}$.

The only assumptions we need to apply in the non-parametric 
approach in our case, are the following:
\begin{enumerate}
\item it is possible to select some area in the region under study relatively 
free of cluster stars, and to determine the frequency function corresponding 
to the VPD of this area. This will provide a 
representation of the field frequency function with a small (or negligible) 
cluster contribution, and
\item the frequency function found this way is representative of the
field frequency function over the whole plate and, specifically, in the area
occupied by the cluster.
\end{enumerate}

The empirical frequency function determined from the VPD corresponding to
the area occupied by the cluster, $\Psi_{c+f}$, is made up of two
contributions: cluster and field.
To differentiate the two populations we
need to estimate the field distribution.
For this purpose, we studied the VPD for the plate area outside a circle
centered on the cluster. The center of the cluster was chosen as the point
of highest spatial density. We did tests with circles of very different
radii, searching a reasonable tradeoff between cleanness (absence 
of a significant amount of cluster members) and signal-to-noise ratio
(working area not too small). The kernel density estimator was 
applied in the VPD to these data, yielding the empirical frequency function,
for a grid with cell size of 0.2 mas~yr$^{-1}$, well below the proper 
motion errors.

\begin{figure}
\begin{center}
\resizebox{7cm}{!}{\includegraphics{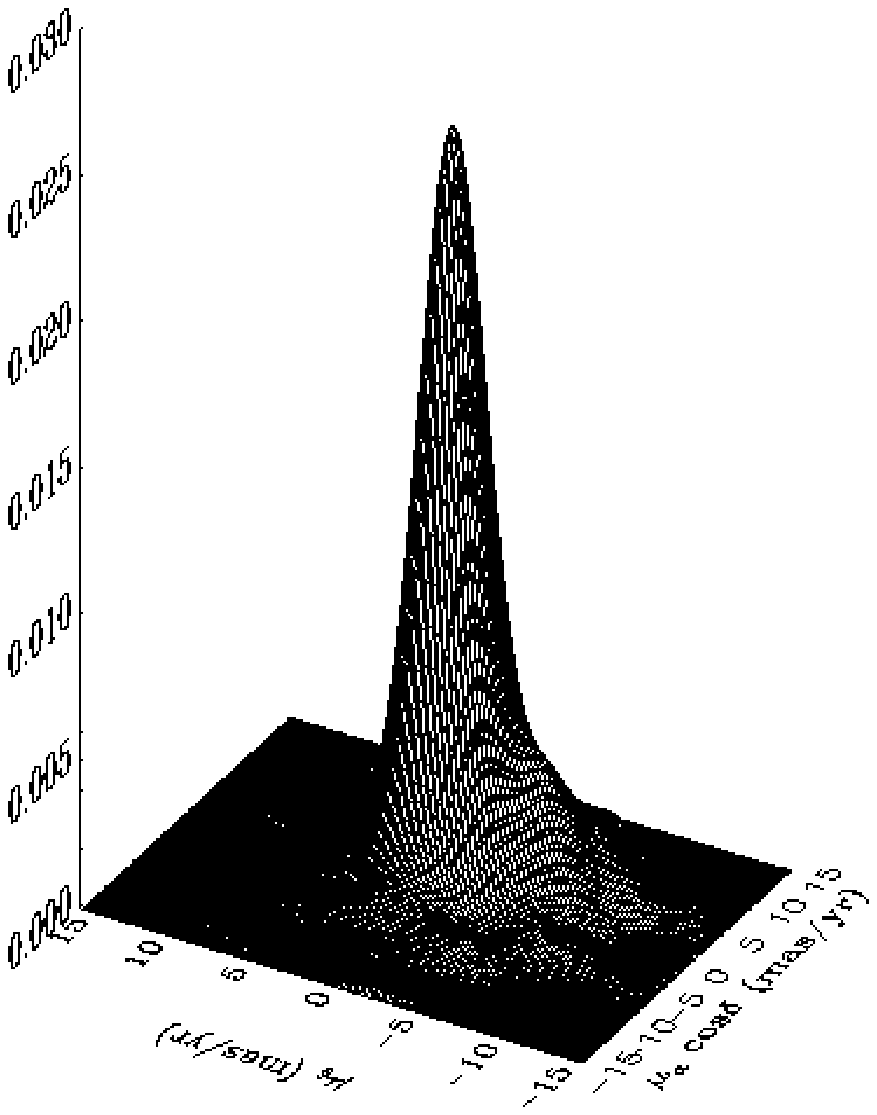}}\\
\resizebox{7cm}{!}{\includegraphics{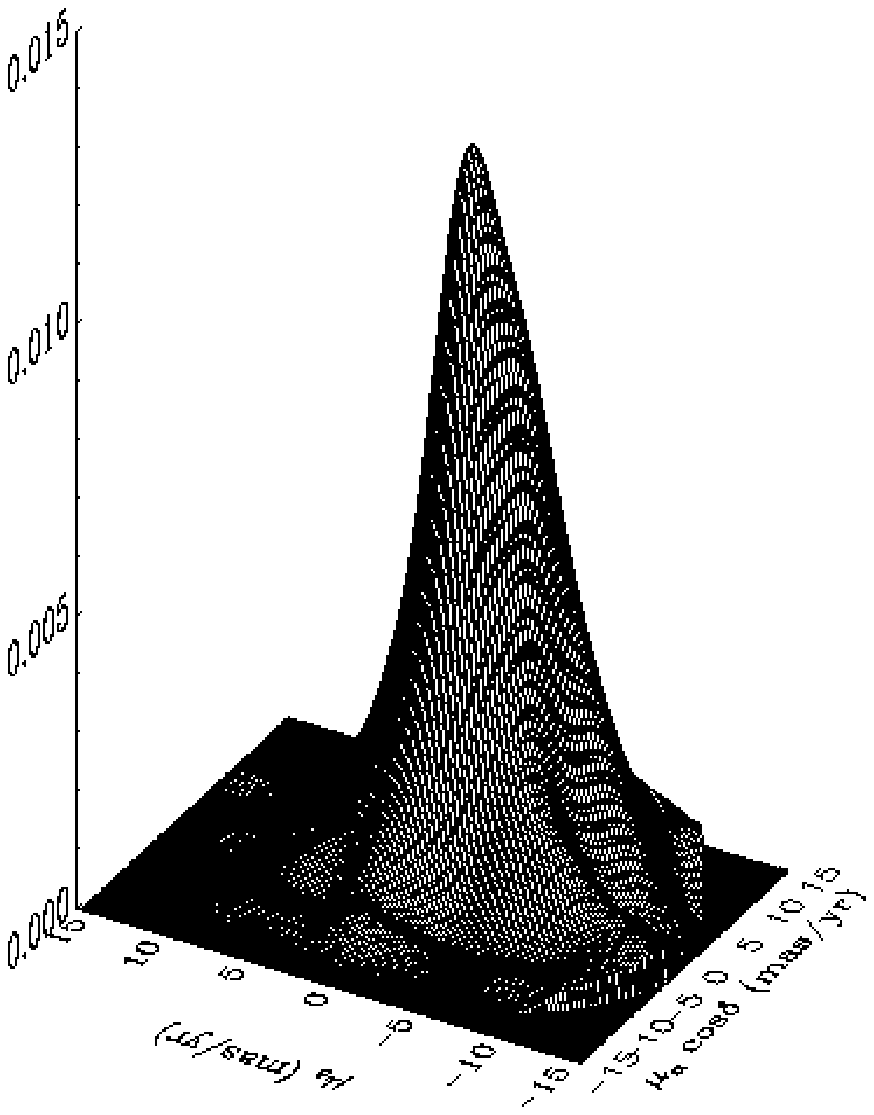}}\\
\resizebox{7cm}{!}{\includegraphics{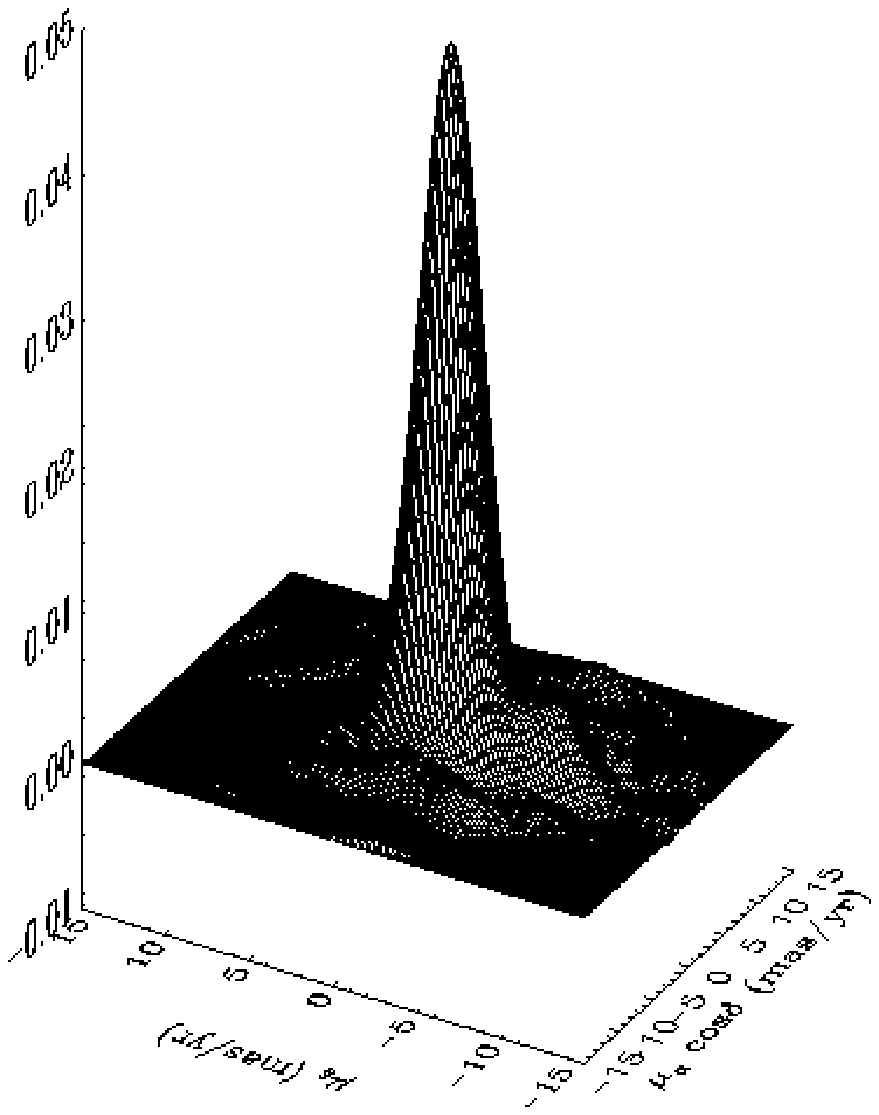}}
\end{center}
\caption{Empirical probability density functions in the kinematic
plane. Top: mixed sample
from the inner circle of 20'. Center: field population from 
outside this circle. Bottom: cluster population of NGC~1817}
\label{nprobo}
\end{figure}

We finally found that the area outside a circle with a radius of 
20$\arcmin$ centered on the cluster yields a clean frequency 
function with low cluster contamination and low noise. This
way we deduce the frequency function representative of the 
field population. The next step was to determine $\Psi_{c+f}$ 
from a plate area centered on the cluster and containing 
both populations (cluster and field).
We found that in our case the inner circle with a radius of 
20$\arcmin$ is well suited for our purposes.
Assuming that the spatial stellar density of the field population
is approximately uniform over the whole area surveyed, we can scale 
the field frequency function previously computed to represent the field
frequency function in the inner circle $\Psi_{f}$ by simply
applying a factor linked to the area.
The cluster empirical frequency function can then be determined as 
$\Psi_{c}$ =$\Psi_{c+f} - \Psi_{f}$.
These empirical frequency functions can be normalized to yield the 
empirical PDFs for the mixed populations (circle), for the field 
(outside the circle) and for the cluster (non-field) population. 
Figure~\ref{nprobo} displays these three functions. 
The probability for a star in a node of the grid being a member
of the cluster is 
$P_{c}$ = ($\Psi_{c+f}$ - $\Psi_{f}$)/$\Psi_{c+f}$.
The empirical tables can then be used to estimate the probability 
of a star being a cluster member according to the probability of 
its nearest node. These
probability tables are then applied to all the stars
in the surveyed area, both inside and outside the circle defined 
to determine the functions.

Of course, the field PDF estimated in the outer area cannot be an absolutely
perfect representation of the true field PDF in the whole area. This 
introduces undesired noise in the frequency function of the cluster.
The negative density values found in several zones obviously lack 
physical meaning. These negative values 
allow us to estimate the typical noise level, $\gamma$, present in the result.
To avoid meaningless probabilities in zones of low density we restricted the
probability calculations to the stars with cluster PDF $\geq$ 3$\gamma$.
The maximum of the cluster PDF is located at 
($\mu_{\alpha}\cos\delta, \mu_{\delta}$) = (0.0$\pm$0.2,$-$0.8$\pm$0.2) mas~yr$^{-1}$.

Like the negative density values, the small local maximum found in the 
cluster PDF around ($\mu_{\alpha}\cos\delta, \mu_{\delta}$) = (0.0,$-$6.0) mas~yr$^{-1}$
 is also due to the fact that the empirical frequency function computed in the outer 
area of the plate does not represent the inner circle field frequency function 
with absolute accuracy. The stars with proper motions in this VPD area are 
spread over the plate and their photometry (when available from 
Paper~III) does not suggest that they correspond to any physical group.


\section{Results and discussion}

\begin{figure}
\begin{center}
\resizebox{\hsize}{!}{\includegraphics{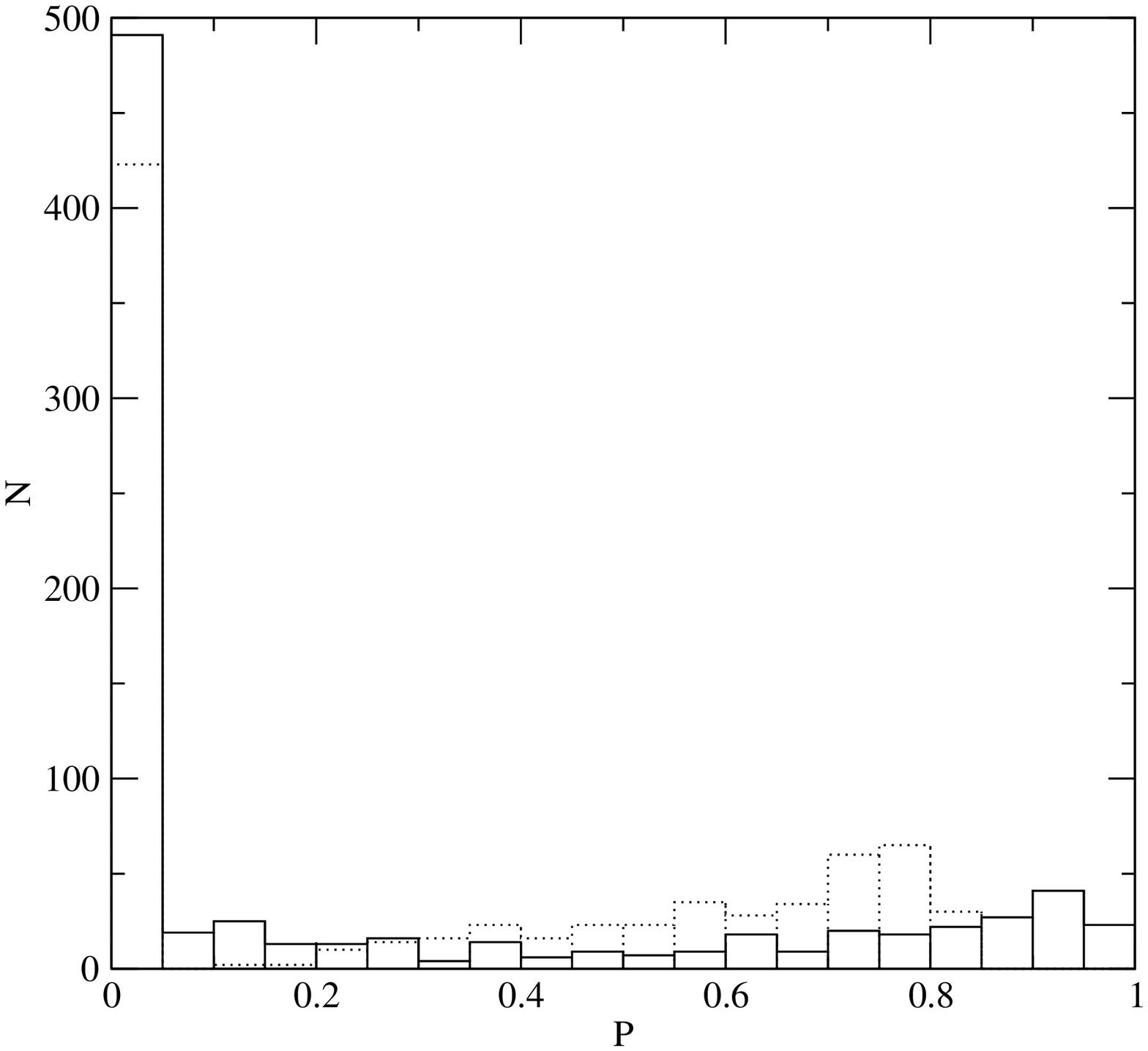}}
\end{center}
\caption{The histogram of cluster membership probability 
of NGC~1817. The solid line gives the results for traditional parametric 
method, while the dotted line corresponds to the non-parametric approach.}
\label{fig7}
\end{figure}

	The results from the non-parametric and parametric approaches 
are in agreement, which indicates the reliability of both methods
in the case of this cluster.
While in the parametric approach we  
need to impose a null internal dispersion (based on the known 
distance of the cluster) for a reliable segregation, 
in the non-parametric approach we are able to differentiate 
the cluster population without the need of any a priori knowledge. 

Furthermore, the parametric approach is quite sensitive to the initial 
values used for the iterations, and special care has to be taken at 
every step to ensure that the final results make sense from a physical 
point of view. In the case of a doubtful number of independent clusters present 
in an area the parametric method can be misleading (as happened in Paper~I) 
and additional information has to be introduced 
explicitly to get the PDFs. On the contrary, as shown by Gal98, if 
there is more than one cluster in a zone (and they show a distinctive 
kinematic behaviour), the non-parametric approach is capable of detecting 
and managing them in a direct and natural manner. In our case, we detected
no sign of a distinct cluster NGC~1807. 

On the other hand, the non-parametric approach does not take into account the
errors of the individual proper motions, 
therefore it does not make any particular distinction between bright or faint 
stars, different epoch spread and so on. 
The FWHM of the empirical cluster PDF provides an estimation of the 
errors of the distribution. We obtained a FWHM of $\sim$4.0$\pm$0.2 
mas~yr$^{-1}$.
If the gaussian dispersion owing to the smoothing 
parameter $h =$1.33 mas~yr$^{-1}$ is taken into account, 
this FWHM corresponds to a mean error in the 
proper motions of 1.5 mas~yr$^{-1}$, 
of the same order as the values given in Sect.~2.2.

    The cluster membership probability histogram (Fig.~\ref{fig7}) 
shows a clear separation between cluster members and field stars
in both approaches: the solid line is the traditional parametric method
while the dotted line is the non-parametric approach. 
But the exact point of deciding which probability means that a star is a 
member has been traditionally left to a usually conservative, but 
subjective, arbitrary decision. 
In our case, the non-parametric approach gives an expected number of
cluster members from the integrated volume of the cluster 
frequency function $\Psi_{c}$ in the VPD areas of high cluster density 
(where $\Psi_{c}> 3\gamma$). 
This integration predicts that the sample contains 135 
cluster members. Sorting the sample in order of decreasing 
non-parametric membership probability, $P_{NP}$, the first 135 stars 
are the most probable cluster members, according to the results of the 
non-parametric technique. The minimum value of the non-parametric 
probability (for the 135-th star) is $P_{NP}=72\%$.

There is no an equivalent rigurous way to decide where to set the 
limit among members and non-members in the list sorted in order of 
decreasing parametric membership probability, $P_P$. But, if we accept 
the size of the cluster predicted by the non-parametric method, 135 
stars, we can consider that the 135 stars of highest $P_P$ are the most 
probable members, according to the results of the parametric 
technique. The minimum value of the parametric probability (for the 
135-th star) is $P_P=74\%$.

With these limiting probabilities ($P_{NP}\geq0.72$; $P_P\geq0.74$), we get 
a 92$\%$ (743 stars) agreement in the segregation yield by the two 
methods. The 67 remaining stars (8$\%$) with contradicting segregation 
should be carefully studied. Discrepancies among the two approaches are 
actually expected due to the statistical nature of the methods themselves.

Thus, we find ourselves with two lists of member candidates that are 
not in complete agreement. To set up a final and unique list, 
and being conservative, we accept as probable members of this cluster 
those stars classified as members by at least one of the two methods. 
This is equivalent to merging both lists (each with 135 stars) and 
eliminating duplicated entries. This way we get a list of 169 probable 
member stars.

\begin{figure*}
\begin{center}
\resizebox{\hsize}{!}{\includegraphics{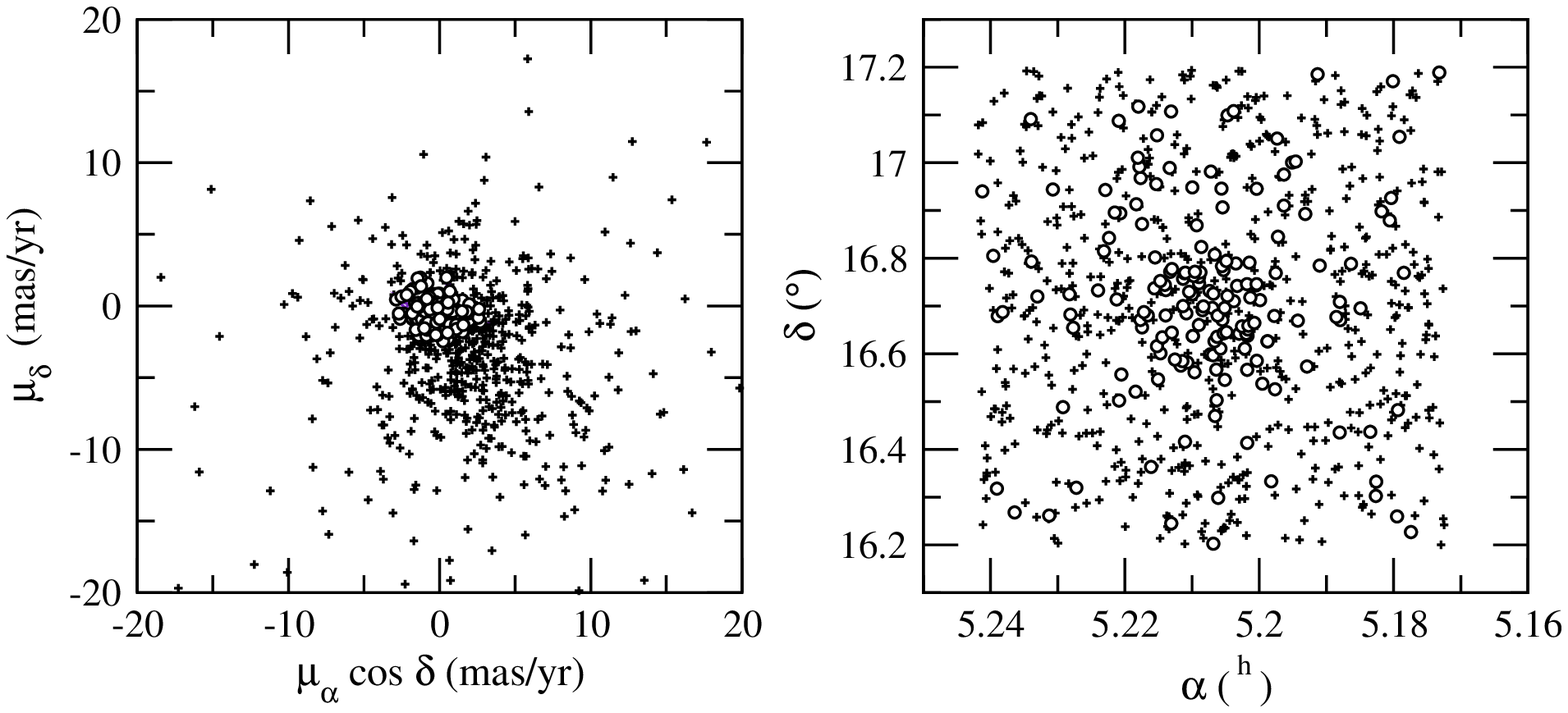}}
\caption{The proper motion vector-point diagram and position
         distribution of stars in NGC~1817 
           (``$\circ$" for members of NGC~1817,
            ``$+$" for field stars)}
\label{mvpr}
\end{center}
\end{figure*}

As in any other cluster membership study based on kinematic information, 
the resulting list of probable members has to be complemented with additional 
information for a cleaner segregation (see Paper~III).

	In Paper~I, 416 stars were considered members of NGC~1817 
($P_1 >$ 0.7), while 14 stars were considered members of NGC~1807 
($P_2 >$ 0.7). Only one of those 14 stars is a member of NGC~1817
according to this study, and the rest belong to the field.   
A detailed comparison of Paper~I with our results for NGC~1817 
shows 113 members in common with the
parametric segregation and 120 in common with the non-parametric one.  
Hence, Paper~I was overestimating the cluster population by as much as
approximately 300 stars that most probably belong to the field.

   Figure~\ref{mvpr} shows the proper motion VPD
   and the sky distribution for all the measured
   stars, where $``\circ"$ denotes a selected member of NGC~1817,
   and all other stars are considered field stars indicated
   by ``$+$".

	A comparison with the 76 stars included in the radial velocity study 
by Mer03 is given 
in Table~\ref{vr}. The radial velocities have errors of $\approx$ 0.5 km~s$^{-1}$.
To quantify the differences we set an agreement 
index $P_{c}$ to 1 if 
the parametric probability, $P_{P}$, agrees with the radial velocity 
segregation, 2 if the non-parametric probability, $P_{NP}$, agrees, 
 3 if both probabilities agree and 0 if none does. 
We find 62 out of 76 stars with $P_{c} >$~0, that is 82$\%$ agreement
with the radial velocities segregation. 
18$\%$ of the disagreement consists of  
10 stars out of 36 (28$\%$) being considered non-members on the basis 
of proper motions 
while only 4 out of 40 (10$\%$) were found to be astrometric 
members while considered non-members on the basis of radial velocities.  
  
If we compare the two methods, the behaviour is rather similar. For the 
parametric method we find a total of 60 stars (79$\%$) whose membership
assignation coincides with the radial velocity criterion, while
for the non-parametric method this amounts to 54 stars (71$\%$).

\addtocounter{table}{1}

	The results show that the two approaches are
similar when the parameters are well established in the parametric method
and when a suitable area free from cluster stars is chosen in the non-parametric
technique. 
But we need to be aware of the risks of the parametric model when there is
more than one cluster or probable cluster. We consider the non-parametric 
approach a good alternative to avoid mathematical artefacts. 

Table~5 lists the results for all 810 stars in the region of the
open cluster: column 1 is the ordinal star number (as in Paper~I,
the numbering system comes from the PDS measuring machine); 
columns 2 and 3
give $\alpha_{\mathrm J2000}$ and $\delta_ {\mathrm J2000}$;
columns 4 and 6 list the respective absolute proper motions 
($\mu_{\alpha}\cos\delta, \mu_{\delta}$); columns 5 and 7 contain the
standard errors of the proper motions; 
column 8 gives the number of plates used to derive proper motions;
column 9 and 10 are the parametric
and non-parametric membership probabilities of stars belonging to NGC~1817 
and column 11 provides the identification number
in the Tycho-2 Catalogue for the stars in common.

   The present results for NGC~1817, based on astrometric data only,
 are complemented with the photometric study of Paper~III. 


\section{Summary}

    Proper motions and their corresponding errors for 810 stars within
    a 1\fdg5 $\times$ 1\fdg5 area in
    the NGC~1817 region were determined from PDS measurements
    of 25 plates with a baseline of 81 years. A comparison with the Tycho-2
    Catalogue shows good agreement and underlines the precision
    of the proper motions derived in this paper. These proper motions are
    then used to determine membership probabilities of the stars in the region.
    By combining parametric and non-parametric approaches,  
    this new membership study leads to a much better segregation of the 
    cluster stars. 
    We obtained a list with 169 probable member stars.

\vspace{3mm}
%

\

\begin{acknowledgements}
    We would like to thank Floor van Leeuwen 
    for his continuous help and 
    valuable comments, as well as all the people at the IoA (Cambridge)
    for a very pleasant stay. L.B-N. gratefully acknowledges financial 
    support from EARA Marie Curie Training Site (EASTARGAL) during her 
    stay at IoA.
    This study was also partially
    supported by the contract No. AYA2003-07736 with MCYT.
    This research has made use of Aladin, developed by CDS, 
Strasbourg, France.

\end{acknowledgements}

\Online

\setcounter{figure}{0}
\setcounter{table}{0}

\begin{figure}
\begin{center}
\resizebox{\hsize}{!}{\includegraphics{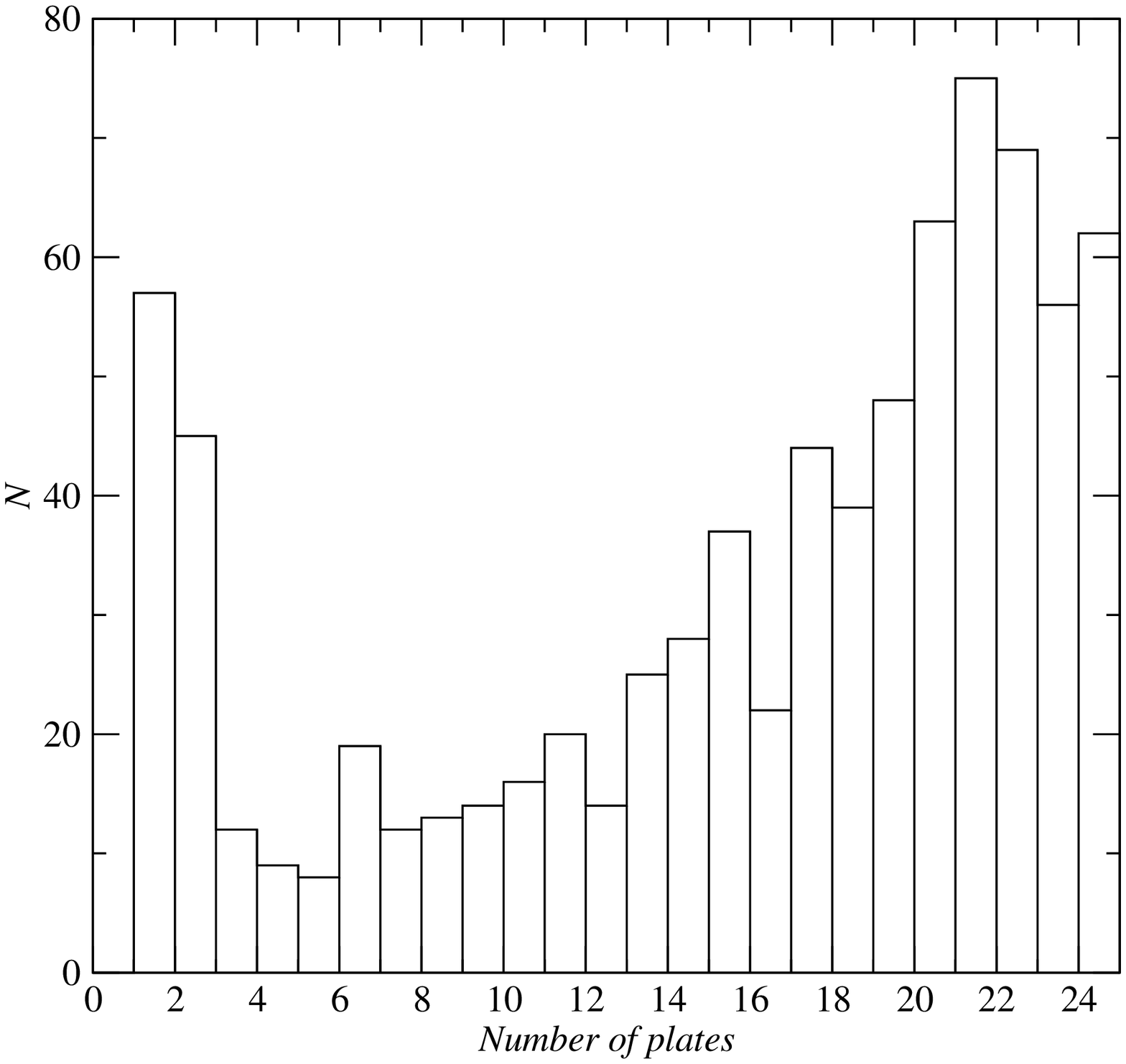}}
\end{center}
\caption{The number of stars ($N$) {\it vs} the number of available plates}
\label{nplpr}
\end{figure}

\begin{table}
\leavevmode
\caption {Shanghai Observatory (Z\^ o-S\`e astrograph) plates used 
in this study. 
}
{\small
\begin{center}
\begin {tabular} {ccccc}
\hline
   Plate  & Epoch & Exp.time & N. of  \\
  id. & (1900+) & min:sec  &  stars \\
\hline

     CL415   &  16.01.31  &  30   &  159 \\
     CL5293  &  30.02.16  &  --   &  660 \\
     CL5291  &  43.02.09  &  --   &  537 \\
     CL5292  &  43.02.22  &  --   &  634 \\
     CL54002 &  54.01.31  &  80   &  684 \\
     CL54003 &  54.02.08  &  90   &  656 \\
     CL61010 &  61.02.02  &  40   &  848 \\
     CL61003 &  61.02.12  &  20   &  602 \\
     CL61002 &  61.02.17  &  30   &  714 \\
     CL81009 &  81.12.24  &  20   &  464 \\
     CL81010 &  81.12.24  &  20   &  508 \\
     CL81011 &  81.12.25  &  15   &  588 \\
     CL81012 &  81.12.25  &  15   &  512 \\
     CL82001 &  82.02.24  &  15   &  701 \\
     CL82002 &  82.02.24  &  15   &  678 \\
     CL82003 &  82.02.24  &  12   &  655 \\
     CL82004 &  82.02.27  & 8:15  &  606 \\
     CL82005 &  82.02.27  & 5:45  &  373 \\
     CL82006 &  82.02.27  &  15   &  651 \\
     P9701   &  97.01.02  &  30   &  470 \\
     P9702   &  97.01.11  &  30   &  426 \\
     P9703   &  97.01.11  &  30   &  478 \\
     P9704   &  97.01.11  &  30   &  168 \\
     P9705   &  97.01.12  &  30   &  587 \\
     P9706   &  97.01.12  &  30   &  540 \\

\hline
\label{plates}
\end{tabular}
\end{center}
}
\end{table}

\begin{table}
\caption{Mean precisions of proper motions as a function of the     
          number of plates in the NGC~1817 region. 
(Units are mas~yr$^{-1}$.) Columns "N. plates" and "$N$" give 
the number of plates and stars, respectively.}
\begin{center}
\begin {tabular} {ccccc}
\hline
 N. plates & $N$ & $\epsilon_{\mu_{\alpha}\cos\delta}$ & $\epsilon_{\mu_{\delta}}$ & $\epsilon_{\mu}$ \\
\hline
 4 &   10 & 1.359 & 1.235 & 1.932 \\
 5 &    7 & 1.593 & 1.455 & 2.279 \\
 6 &    7 & 1.624 & 1.338 & 2.249 \\
 7 &   14 & 1.790 & 1.520 & 2.403 \\
 8 &   10 & 1.996 & 1.390 & 2.464 \\
 9 &   12 & 1.720 & 1.765 & 2.529 \\
10 &   13 & 1.684 & 1.429 & 2.249 \\
11 &   13 & 2.012 & 1.658 & 2.655 \\
12 &   20 & 1.782 & 1.420 & 2.351 \\
13 &   14 & 1.611 & 1.290 & 2.136 \\
14 &   24 & 1.635 & 1.299 & 2.134 \\
15 &   28 & 1.515 & 1.486 & 2.166 \\
16 &   37 & 1.531 & 1.095 & 1.933 \\
17 &   21 & 1.514 & 1.206 & 1.982 \\
18 &   44 & 1.172 & 1.023 & 1.591 \\
19 &   38 & 1.228 & 0.950 & 1.592 \\
20 &   48 & 1.242 & 1.056 & 1.667 \\
21 &   63 & 1.076 & 0.829 & 1.400 \\
22 &   75 & 0.842 & 0.720 & 1.131 \\
23 &   69 & 0.940 & 0.677 & 1.179 \\
24 &   56 & 0.630 & 0.559 & 0.853 \\
25 &   62 & 0.471 & 0.429 & 0.646 \\
$>3$ &685 & 1.162 & 0.958 & 1.545 \\
\hline
\end {tabular}
\end{center}
\label{error}
\end {table}

\addtocounter{table}{1}
\begin{table*}
\caption {The cross-identification of stars in common
         with the radial velocities analysis by Mermilliod et al.\ 
(\cite{Mermi}) and
the comparison of its membership for parametric ($P_P$) and 
non-parametric ($P_{NP}$) results. See text for explanation of
the agreement index $P_{c}$. }
\begin {tabular} {cccccccccccc}
\hline
 Mer03 & Table~5 & $P_{P}$ & $P_{NP}$ & $P_{c}$ & $P_{Vr}$ & 
 Mer03 & Table~5 & $P_{P}$ & $P_{NP}$ & $P_{c}$ & $P_{Vr}$  \\
\hline
    8 & 557 & 0.40 & 0.65  & 0 & M &    90 & 554 &  0.00 & 0.47 & 3 & NM \\
   12 & 562 & 0.88 & 0.77  & 3 & M &   103 & 580 &  0.01 & 0.28 & 3 & NM \\
   19 & 358 & 0.92 & 0.73  & 3 & M &   138 & 322 &  0.57 & 0.62 & 3 & NM \\
   22 & 351 & 0.70 & 0.66  & 0 & M &   155 & 519 &  0.08 & 0.78 & 1 & NM \\
   30 & 334 & 0.65 & 0.63  & 0 & M &   161 & 527 &  0.00 & 0.31 & 3 & NM \\
   40 & 541 & 0.85 & 0.80  & 3 & M &   187 & 394 &  0.00 & 0.00 & 3 & NM \\
   44 & 546 & 0.98 & 0.78  & 3 & M &   269 & 521 &  0.00 & 0.00 & 3 & NM \\ 
   56 & 379 & 0.96 & 0.79  & 3 & M &   531 & 432 &  0.00 & 0.35 & 3 & NM \\ 
   71 & 317 & 0.58 & 0.60  & 0 & M &   536 & 448 &  0.00 & 0.00 & 3 & NM \\ 
   72 & 318 & 0.73 & 0.81  & 2 & M &   571 & 610 &  0.81 & 0.81 & 0 & NM \\ 
   73 & 528 & 0.95 & 0.81  & 3 & M &   598 & 696 &  0.00 & 0.00 & 3 & NM \\ 
   79 & 543 & 0.94 & 0.77  & 3 & M &   621 & 768 &  0.00 & 0.00 & 3 & NM \\ 
   81 & 542 & 0.79 & 0.77  & 3 & M &  1081 & 885 &  0.00 & 0.00 & 3 & NM \\ 
  121 & 367 & 0.94 & 0.71  & 1 & M &  1082 & 886 &  0.00 & 0.78 & 1 & NM \\ 
  127 & 339 & 0.86 & 0.82  & 3 & M &  1083 & 896 &  0.00 & 0.00 & 3 & NM \\ 
  164 & 532 & 0.98 & 0.80  & 3 & M &  1095 & 870 &  0.88 & 0.77 & 0 & NM \\ 
  177 & 577 & 0.81 & 0.63  & 1 & M &  1096 & 863 &  0.00 & 0.34 & 3 & NM \\ 
  211 & 331 & 0.55 & 0.80  & 2 & M &  1112 & 733 &  0.72 & 0.62 & 3 & NM \\ 
  212 & 321 & 0.90 & 0.74  & 3 & M &  1153 & 815 &  0.00 & 0.23 & 3 & NM \\ 
  244 & 556 & 0.92 & 0.75  & 3 & M &  1161 & 811 &  0.00 & 0.00 & 3 & NM \\ 
  185 & 395 & 0.93 & 0.81  & 3 & M &  1194 & 674 &  0.00 & 0.00 & 3 & NM \\ 
  206 & 343 & 0.91 & 0.78  & 3 & M &  1197 & 672 &  0.00 & 0.30 & 3 & NM \\ 
 1114 & 735 & 0.79 & 0.70  & 1 & M &  1246 & 468 &  0.00 & 0.00 & 3 & NM \\ 
 1117 & 600 & 0.97 & 0.74  & 3 & M &  1248 & 467 &  0.15 & 0.54 & 3 & NM \\  
 1135 & 714 & 0.90 & 0.82  & 3 & M &  1252 & 660 &  0.00 & 0.00 & 3 & NM \\ 
 1152 & 816 & 0.00 & 0.35  & 0 & M &  1254 & 648 &  0.00 & 0.44 & 3 & NM? \\ 
 1208 & 689 & 0.03 & 0.28  & 0 & M &  1267 & 237 &  0.00 & 0.00 & 3 & NM \\ 
 1265 & 463 & 0.19 & 0.51  & 0 & M? &  1273 & 112 &  0.00 & 0.00 & 3 & NM \\ 
 1292 & 605 & 0.87 & 0.65  & 1 & M &  1297 & 430 &  0.94 & 0.73 & 0 & NM? \\ 
 1408 & 301 & 0.97 & 0.76  & 3 & M &  1302 & 423 &  0.00 & 0.77 & 1 & NM \\ 
 1412 & 150 & 0.93 & 0.79  & 3 & M &  1314 & 206 &  0.00 & 0.00 & 3 & NM \\ 
 1420 & 160 & 0.75 & 0.75  & 3 & M &  1316 & 214 &  0.00 & 0.21 & 3 & NM \\ 
 1433 & 180 & 0.97 & 0.74  & 3 & M &  1406 & 304 &  0.93 & 0.78 & 0 & NM \\ 
 1456 & 292 & 0.69 & 0.68  & 0 & M &  1418 &  56 &  0.00 & 0.45 & 3 & NM \\ 
 1459 & 296 & 0.03 & 0.66  & 0 & M &  1424 &   3 &  0.00 & 0.27 & 3 & NM \\ 
 1574 &  65 & 0.05 & 0.52  & 0 & M &  1455 & 502 &  0.40 & 0.82 & 1 & NM \\  
      &     &      &   &   &   &  1467 & 493 &  0.00 & 0.00 & 3 & NM \\ 
      &     &      &   &   &   &  1502 & 122 &  0.65 & 0.48 & 3 & NM? \\ 
      &     &      &   &   &   &  1557 &  91 &  0.00 & 0.00 & 3 & NM \\ 
      &     &      &   &   &   &  1718 & 162 &  0.00 & 0.00 & 3 & NM? \\ 
\hline
\label{vr}
\end{tabular}
\end{table*}

\end{document}